\documentstyle[prb,aps,twocolumn,floats]{revtex}

\begin{document}
\input epsf
\draft
\wideabs{
\title {Thermal conductivity and specific heat of the linear chain cuprate 
Sr$_{2}$CuO$_{3}$:\\  
Evidence for thermal transport via spinons
}

\author {A. V. Sologubenko, E. Felder,  K. Giann\`{o}, H. R. Ott}
\address{Laboratorium f\"ur Festk\"orperphysik, ETH H\"onggerberg, 
CH-8093 Z\"urich, Switzerland}

\author{A. Vietkine, A. Revcolevschi}
\address{Laboratoire de Physico-Chimie des Solides, Universit\'e Paris-Sud, 
91405 Orsay, France}

\date{\today}
\maketitle

\begin{abstract}
We report measurements of the specific heat and the thermal 
conductivity of the model Heisenberg spin-1/2 chain cuprate 
Sr$_{2}$CuO$_{3}$ at low temperatures. In addition to a nearly isotropic 
phonon heat transport, we find a quasi one-dimensional 
excess thermal conductivity 
along the chain direction, most likely associated with spin excitations 
(spinons). The spinon energy current is limited mainly by scattering 
on defects and phonons. 
Analyzing the specific heat data, the intrachain magnetic exchange 
$J/k_{B}$ is estimated to be $\simeq$2650~K.  
    
\end{abstract}
\pacs{PACS numbers: 66.70.+f, 
                    75.40.Gb, 
		    74.72.Jt  
		    }
}

There is a considerable theoretical interest in one-dimensional (1D) 
Heisenberg spin-1/2 systems because they exhibit a number of 
properties that are entirely dominated by quantum-mechanical behavior 
and have no analogues in three-dimensional systems. 
In particular, it has been shown that the Heisenberg $S$=1/2 chain 
represents an integrable system characterized by a macroscopic number 
of conservation laws.\cite{Zotos} One important conserved quantity is 
the energy current,\cite{Zotos,Niemeijer} implying an ideal (infinite) thermal 
conductivity along the chains at nonzero temperatures, if 
perturbations from impurities, phonons, or an interchain coupling,
which always lead to non-integrable models, are negligible.
It is an open question, to what extent a real material may be
regarded as an ideal 
integrable system. Probably, the most obvious evidence for the predicted anomalous heat 
transport is the recent observation of an unusually high quasi-1D 
magnon thermal conductivity 
in the series (Sr,Ca,La)$_{14}$Cu$_{24}$O$_{41}$.\cite{Sologubenko,Kudo} 
The structure of these materials 
contains two building blocks with 1D character, namely  CuO$_{2}$ chains and Cu$_{2}$O$_{3}$ 
ladders, both oriented along the same direction.
Unfortunately, the dimerisation within the chains  and  a non-negligible 
interchain interaction in this system
complicate the analysis of the observed features in terms of an integrable 
model.

\begin{figure}[t]
 \begin{center}
  \leavevmode
  \epsfxsize=0.9\columnwidth \epsfbox {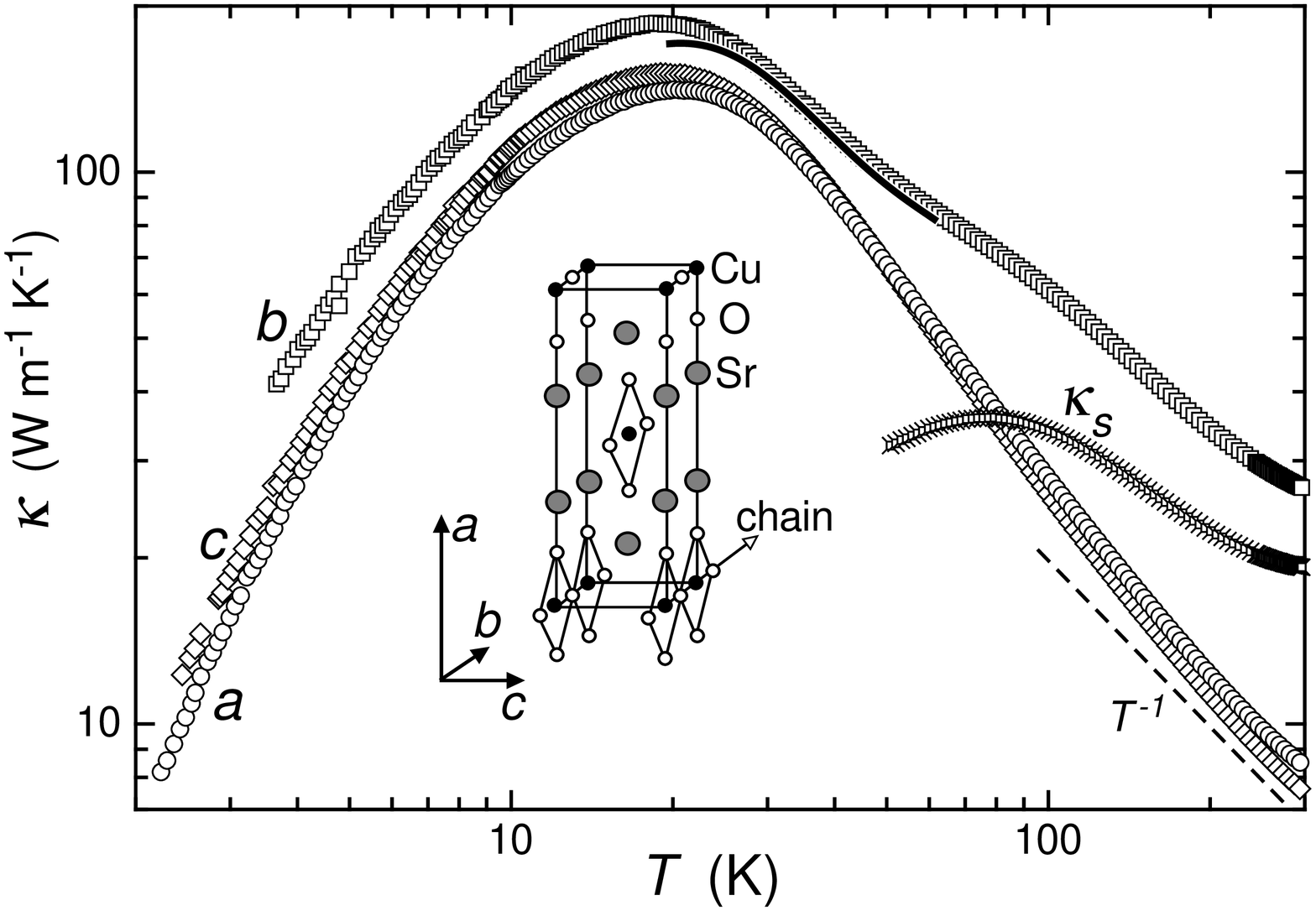}
  \caption{
  Temperature dependence of the thermal conductivity of 
Sr$_{2}$CuO$_{3}$ along the $a$, $b$, and $c$ axes.
$\kappa_{s}$ is the calculated spinon thermal conductivity along the 
$b$ axis.
The solid line is the estimated sum of spinon and phonon thermal 
conductivities assuming that the spinon mean free path is equal to the 
distance between bond-defects (see text).  
The schematic crystal structure is shown in the inset.
  }
  \label{Lamlglg}
 \end{center}
\end{figure}

In this work, we have searched for anomalies in the thermal 
transport of Sr$_{2}$CuO$_{3}$, which is often considered as 
the best physical realization of the 1D Heisenberg $S$=1/2 model. 
The crystal structure of Sr$_{2}$CuO$_{3}$ contains chains formed by 
CuO$_{4}$ squares sharing oxygen corners.\cite{Teske} The chains run along the 
$b$ axis and, as shown in the inset of Fig.~\ref{Lamlglg}, 
the  CuO$_{4}$ squares lie in the $ab$ plane. 
The intrachain exchange interaction between neighboring Cu$^{2+}$ ions connected via
180$^\circ$ Cu-O-Cu bonds, measured as  $J/k_{B}$, is between 
2150 and 3000~K.\cite{Ami,Suzuura,Motoyama,Johnston} 
The ratio $k_{B} T_{N}/J$, where $T_{N}$ is the 3D N\'{e}el temperature,  
is as small as $2\times 10^{-3}$, reflecting an extremely small 
ratio $J'/J$,  $J'$ representing the interchain interaction.

Our observations indicate an excess thermal conductivity along the chain direction, 
provided by quasi-1D spin excitation (spinons). 
According to our analysis presented below, its magnitude  is limited by scattering of 
spinons on defects and phonons. We find no evidence for a mutual scattering between spin 
excitations and hence it seems to be absent or at least negligibly small, in agreement with 
theoretical predictions for integrable models.

The specimens used in these experiments were cut from a single 
crystal that had been grown by the traveling 
solvent floating zone method. The details of crystal growth and 
structural characterization are described elsewhere.\cite{Revc97} 
For thermal transport measurements, three rectangular-bar-shaped samples 
of typical dimensions 
$2.5 \times 1 \times 1$ mm$^{3}$ with the longest dimension parallel 
to either the $a$, $b$ or $c$ axis were prepared.
Two additional samples, \#1 and \#2, cut
from the same piece, were used
for specific heat measurements. The thermal conductivity 
was measured using a conventional steady-state 
method as described in Ref.~\onlinecite{Sologubenko}. A standard 
relaxation technique was employed for the specific heat measurements.   
The magnetic susceptibility $\chi$ was measured 
with a commercial SQUID-magnetometer. 

Small amounts of excess oxygen are known to be present in as-grown 
crystals of Sr$_{2}$CuO$_{3}$,
giving rise to a Curie-Weiss term in the temperature dependence of the
magnetic susceptibility\cite{Ami} due to uncompensated Cu $S$=1/2 spins.
In order to study the influence of excess oxygen, we
annealed sample \#2 
at 870$^{\circ}$C for 72~h under argon atmosphere, 
as described in Ref.~\onlinecite{Motoyama}.
From the results of our measurements of $\chi(T)$, the 
ratio of the number of residual spin-1/2 impurities to the total 
number of  Cu ions was estimated to be 1.8$\times$10$^{-4}$ for the unannealed sample \#1 
and 6$\times$10$^{-5}$ for the annealed sample \#2.

\begin{figure}[t]
 \begin{center}
  \leavevmode
  \epsfxsize=0.9\columnwidth \epsfbox {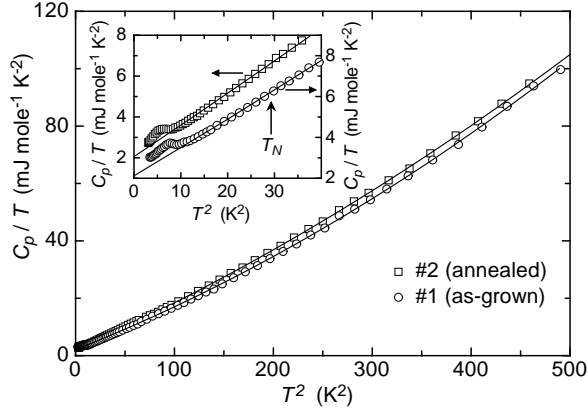}
  \caption{
  Specific heat of Sr$_{2}$CuO$_{3}$ as $C_{p}/T$ versus $T^{2}$. 
  The solid lines represent the fit to Eq.~(\ref{eCp}).  
The inset shows the data at low temperatures. The value of 
  $T_{N}$ denoted by an arrow is from Ref.~18.
  }
  \label{Cp}
 \end{center}
\end{figure}

The results of the specific heat ($C_{p}$) measurements 
in the temperature 
range between 1.5 and 22~K are presented in Fig.~\ref{Cp} as a plot 
of $C_{p}/T$ versus $T^{2}$. 
The solid lines in Fig.~\ref{Cp} are fits to the data above 4~K using 
the approximation
\begin{equation}\label{eCp}
C_{p} = \gamma T + \beta T^{3} + \delta T^{5}.
\end{equation}
The parameter values are 
$\gamma$=2.12 $\times$10$^{-3}$ J~mole$^{-1}$~K$^{-2}$, 
$\beta$=1.359 $\times$10$^{-4}$ J~mole$^{-1}$~K$^{-4}$, and
$\delta$=1.310 $\times$10$^{-7}$ J~mole$^{-1}$~K$^{-6}$
for sample \#1 and 
$\gamma$=2.06 $\times$10$^{-3}$ J~mole$^{-1}$~K$^{-2}$, 
$\beta$=1.531 $\times$10$^{-4}$ J~mole$^{-1}$~K$^{-4}$, and
$\delta$=1.310 $\times$10$^{-7}$ J~mole$^{-1}$~K$^{-6}$
for sample \#2. 
The sum $\beta T^{3} + \delta T^{5}$ in Eq.~(\ref{eCp}) is a 
common low-temperature approximation for the lattice specific heat.
The fit values of the
parameter $\beta$ result in  values of 
the Debye temperature of $\Theta_{D}$=441$\pm$10~K and 424$\pm$10~K for samples 
\#1 and \#2, respectively.
Since Sr$_{2}$CuO$_{3}$ is an insulator, 
the linear in $T$ contribution is not due to itinerant electrons
but is ascribed to spin degrees of freedom. 

The elementary excitations of a 1D Heisenberg spin one-half
antiferromagnetic system are not $S$=1 magnons but $S$=1/2 topological excitations,\cite{Faddeev} 
now commonly called "spinons". The applicability of the spinon model has been demonstrated experimentally
for several $S$=1/2 chain systems, including Sr$_{2}$CuO$_{3}$.\cite{Misochko,Fujisawa}  
The corresponding specific heat at $T \ll J/k_{B}$ is given by\cite{Takahashi,Haldane,McRae}
\begin{equation}\label{eCmag}
C_{s} =  \frac{2Nk_{B}^{2}}{3J}T,
\end{equation}
where $N$ is the number of magnetic ions in the system.
The fit values of the parameter $\gamma$ give $J/k_{B}$=2620$\pm$100~K and 
2690$\pm$100~K for samples \#1 and \#2, respectively. 
This may be compared with $J/k_{B}\simeq$2200~K deduced from magnetic 
susceptibility data\cite{Motoyama,Johnston} and a somewhat larger 
value of $J/k_{B}$=2850~K, as obtained from the analysis of an
optical absorption spectrum.\cite{Suzuura,Lorenzana} 
The cited susceptibility measurements covered a wide 
temperature range between 5 and 800~K, whereas the absorption 
spectrum presented in Ref.~\onlinecite{Suzuura} 
has been recorded at low temperatures (32~K).  
Hence the discrepancy between the values of $J$ may be ascribed to 
its possible decrease with increasing 
temperature.\cite{Takigawa97_56} Our rather large low-temperature values of $J$ are 
compatible with this suggestion.

The inset of Fig.~\ref{Cp} reveals anomalies of the specific heat 
with onsets below approximately 3.5~K for both samples, indicating some 
sort of phase transition. 
For the annealed sample the anomaly is  
shifted to lower temperatures with respect to the peak for the as-grown 
sample. Both anomalies occur at lower temperatures than the  
N\'{e}el temperatures $T_{N}$ found by $\mu$SR\cite{Keren} 
($4.15 < T_{N} < 6$~K) and neutron scattering\cite{Kojima} 
($T_{N}=5.4$~K) measurements, respectively. 
One possible explanation for this disagreement is that, 
besides the transition to an antiferromagnetically (AFM) ordered 
state at $T_{N} = 5.4$~K, not reflected in $C_{p}(T)$, 
there is another transition at $T_{c} < 3.5$~K.  
Recently, two subsequent magnetic phase transitions at 
$T_{c1}=5.0$~K and $T_{c2}=1.5$~K were observed for 
SrCuO$_{2}$, containing similar spin-1/2 chains but assembled pairwise in 
arrays of zigzag chains.\cite{Zaliznyak}  
The more likely possibility that is consistent with our observations 
is that our anomalies reflect the AFM transition 
reported in Refs.~\onlinecite{Keren} and \onlinecite{Kojima}, but now
shifted to lower temperatures because of a smaller number of 
impurities in the samples. It has been shown that nonmagnetic 
impurities interrupting the spin-1/2 chains enhance staggered spin-spin 
correlations,\cite{EggertImp} a common 
feature of various low-dimensional Heisenberg spin systems.\cite{Laukamp}      
A convincing manifestation of this feature is the stabilisation 
of the long-range AFM order by replacing Cu$^{2+}$ with non-magnetic 
ions in the spin-ladder system 
SrCu$_{2}$O$_{3}$ (Ref.~\onlinecite{Azuma}) and the spin-Peierls system 
CuGeO$_{3}$ (Refs.~\onlinecite{Oseroff,Hase}). Recently, a field-induced staggered 
magnetisation near impurities was also observed in 
Sr$_{2}$CuO$_{3}$ by NMR measurements.\cite{Takigawa97_55} Therefore, 
it seems quite likely that the experimentally observed N\'{e}el temperatures
are always enhanced via the influence of impurities and exceed the 
value that is given by the ratio $J'/J$ itself. Indeed, for Sr$_{2}$CuO$_{3}$ 
calculations considering only dipolar interchain 
coupling\cite{Ami} yield a value of $T_{N}$  as low 
as 0.028~K.

The results of the thermal conductivity $\kappa$ along the $a$, $b$, and $c$ axes 
are presented in Fig.~\ref{Lamlglg}. 
For both directions perpendicular to the chain direction, $\kappa(T)$ shows a peak at 
$T_{\rm max} \sim 20$~K and a decrease with increasing temperature, 
tending to a $T^{-1}$ variation above $T \geq 200$~K.
This behavior is typical for phonon thermal 
transport.\cite{Berman} The thermal conductivity along 
the chain direction, $\kappa_{b}$, 
exhibits the same 
temperature dependence at $T \leq T_{\rm max}$ but obviously 
not so at higher temperatures. 
We suggest that this difference 
is caused by an additional quasi-1D heat transport 
along the chain direction, provided by spin excitations.   
For insulators the phonon-phonon scattering mechanism leads to 
$\kappa \propto T^{-n}$ ($n \sim 1$) 
at $T \geq \Theta_{D}$. For layered structures, such as  
Sr$_{2}$CuO$_{3}$, where the layers are perpendicular to the $a$ axis, 
one may expect that in this temperature region the  ratio between an in-plane and 
the out-of-plane phonon conductivity is larger than the anisotropy of 
$\kappa(T)$ along two different in-plane directions.\cite{Ren}
Since the difference between $\kappa_{c}$ (in-plane) and $\kappa_{a}$ 
(out-of-plane) is 
very small, the difference between $\kappa_{b}$ and $\kappa_{c}$ (both 
in-plane) should even be smaller. This argument
is, of course, not valid for the temperature region near and below 
$T_{\rm max}$, where the influence of sample boundaries and various 
defects is important and, therefore, the behavior of $\kappa(T)$ is,
to some extent, sample-dependent.  Our fitting of $\kappa_{c}(T)$ 
and $\kappa_{a}(T)$ of Sr$_{2}$CuO$_{3}$, employing the Debye model of phonon thermal 
conductivity in a similar way as described 
in Ref.~\onlinecite{Sologubenko}, has shown that 
the influence of boundary scattering on the thermal conductivity is 
negligible at $T \geq 50$~K. The same calculation indicates that an 
enhanced phonon contribution to $\kappa_{b}$ below 50~K may mostly be 
traced back to an enhanced phonon mean free path which is limited by 
boundary and defect scattering.  
Based on all these arguments presented 
above, we make the crucial assumption 
that the phonon thermal conductivity at $T \geq 50$~K is almost 
isotropic. 

In order to single-out the heat transport due to spin excitations, 
the phonon thermal 
conductivity, averaged over the $a$- and $c$-directions at $T \geq 
50$~K, was subtracted 
from the experimental data of $\kappa_{b}$. The resulting spin part 
$\kappa_{s}$ is also plotted in Fig.~\ref{Lamlglg}. 
In a first approximation, valid for any system of quasiparticles,
the 1D thermal 
conductivity is given by the simple kinetic expression 
$\kappa_{s} = C_{s} v_{s} l_{s}$ where $C_{s}$ is the specific
heat, $v_{s}$ is the velocity, and $l_{s}$ is the mean free 
path of the spin excitations. The velocity of spinons is\cite{Faddeev}
$v_{s} = J a \pi /2\hbar$, where $a$ is the distance between the
spins along the chain direction. 
Since $T \ll J/k_{B}$ still holds, Eq.~(\ref{eCmag}) for the specific heat
is valid, and thus the thermal 
conductivity of spinons is given by the simple equation
\begin{equation}\label{eLamMag}
\kappa_{s} =  N_{s} a \frac{k_{B}^{2}\pi}{3 \hbar} l_{s} T,
\end{equation}
where $N_{s}$ is the number of spins per unit volume.
Eq.~(\ref{eLamMag}) has been shown to also be valid for 1D magnon 
systems.\cite{Huber}

\begin{figure}[t]
 \begin{center}
  \leavevmode
  \epsfxsize=0.9\columnwidth \epsfbox {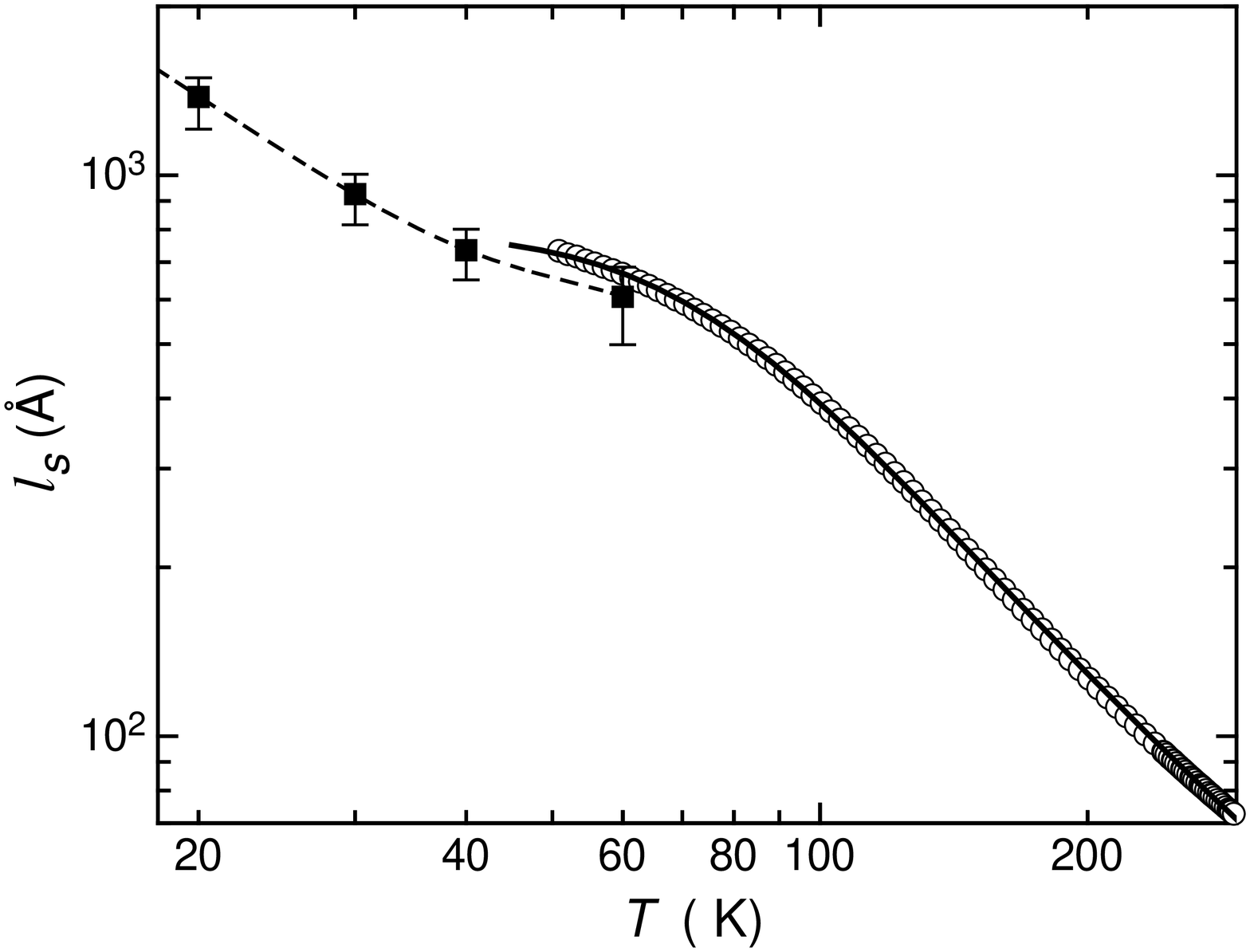}
  \caption{
  Spinon mean free path for Sr$_{2}$CuO$_{3}$. The solid line is a 
  fit to Eq.~(\ref{eSpinMFP}). The solid squares  represent the
  distance between two neighboring bond-defects (from 
  Ref.~30) and the dashed line is a polynomial fit to 
  these data. 
  }
  \label{MagMFP}
 \end{center}
\end{figure}

We calculated $l_{s}$ using Eq.~(\ref{eLamMag}) and taking into 
account small (maximum 6.5\% at 300~K) 
deviations\cite{Johnston2000} of $C_{s}(T)$ from 
linearity. 
The calculated spinon mean free path is shown in Fig.~\ref{MagMFP}.
It tends to reach a constant value at low temperatures 
and decreases with increasing temperature. 
Assuming that the different scattering mechanisms act independently, 
the inverse total mean free 
path of spinons $l_{s}^{-1}$ may be written as a sum $\sum l_{s,i}^{-1}$ 
of the inverse mean free paths 
produced by each scattering mechanism. 
It turns out that
\begin{equation}\label{eSpinMFP}
l_{s}^{-1} = A T \exp(-T^{*}/T) +  L^{-1},
\end{equation}
with the parameters 
$A$=8.2$\times10^{5}$~m$^{-1}$K$^{-1}$, $T^{*}$=186~K, and $L$=7.86$\times10^{-8}$~m,  
reproduces our results in the whole temperature region between 
50 and 300~K with high accuracy, as may be seen in Fig.~\ref{MagMFP}.
If the first term on the right-hand side of Eq.~(\ref{eSpinMFP}) 
is to represent Umklapp processes, spinon-spinon scattering may be 
ruled out because this case would require $T^{*}$ to be of the order 
of $J/k_{B}\sim$2600~K. Since  $T^{*}$ is close to 
$\Theta_{D}/2$=220~K, spinon-phonon Umklapp processes are the most 
likely choice for this contribution.

The second and constant term in Eq.~(\ref{eSpinMFP}) is attributed to 
scattering of spin excitations on defects interrupting the Cu-O 
chains. In that case  the value of the parameter $L$ is a measure for 
the mean distance between the defects. 
Surprisingly, this distance is much shorter than the average distance
of 2$\times$10$^{4}$~${\rm \AA}$ between 
$S$=1/2 impurities,  estimated from the 
Curie-Weiss-type term of the magnetic susceptibility. 
A similar discrepancy has also been encountered in the interpretation 
of NMR measurements on Sr$_{2}$CuO$_{3}$.\cite{Takigawa97_55} 
Recently, in order to explain some peculiar features of NMR spectra 
of Sr$_{2}$CuO$_{3}$, Boucher and Takigawa\cite{Boucher} 
introduced the concept of mobile ``bond-defects'', 
which are not related to interstitial excess oxygen.
The calculated  mean 
distance between two neighboring bond-defects, 
consistent with the NMR data between 20 and 60~K,\cite{Boucher}
is shown in Fig.~\ref{MagMFP}.
We note a  reasonably 
good overlap with our data of $l_{s}$ at $T=60$~K.
The model of Boucher and Takigawa predicts that 
at lower temperatures the interaction between 
defects is important and the number of bond-defects decreases 
with increasing temperature. 
If our parameter $L$ in Eq.~(\ref{eSpinMFP}) indeed represents the distance 
between bond-defects, it ought to be temperature-dependent
below 50~K.
Unfortunately, a quantitative check of this conjecture is difficult because 
of  the uncertain
subtraction of the phonon background at $T \leq 50$~K. 
However, the idea that the  bond-defects are the main source of spinon scattering at 
low temperatures is, at least qualitatively, consistent with the temperature 
dependence and the anisotropy of the thermal conductivity also below 50~K. 
To demonstrate this, we calculated the total thermal conductivity 
($\kappa_{\rm ph} + \kappa_{s}$) along the $b$ axis  at temperatures between 20 and 60~K
assuming, first that $\kappa_{\rm ph}$ is isotropic also in this 
temperature range and equal to the 
average of $\kappa_{c}$ and $\kappa_{a}$ and, second that $l_{s}$ is equal 
to the distance between neighboring bond-defects given in Ref.~\onlinecite{Boucher} (the 
dashed line in Fig.~\ref{MagMFP}). The resulting temperature 
dependence of $\kappa_{b}$ shown by the solid 
line in Fig.~\ref{Lamlglg} is in qualitative agreement with the 
experiment.\cite{Comment1}

Concluding this paper, we return to 
the question whether the present results are relevant vis \`{a} vis of
integrable models.  
We argue that a sizeable quasi-1D 
thermal transport mediated by spin excitations does exist in Sr$_{2}$CuO$_{3}$. 
Its magnitude is not exceptional but 
the scatterers, i.e., defects and phonons, limiting the mean free path 
of spin excitations are extrinsic to the magnetic system. 
Our analysis indicates the absence or negligibly small influence of 
spinon-spinon scattering  on the thermal 
conductivity, 
in agreement with the predictions made for integrable models.\cite{Zotos}
Our results imply that  a dissipationless energy current, 
expected for systems that fulfill the assumptions of  
the integrable Heisenberg 1D $S$=1/2 model, is not 
robust if perturbations like defects and lattice excitations interfere. 
The very high value of the magnetic exchange interaction within the 
chains and the low temperature  magnetic phase transition, 
identified via our specific  heat measurements, 
confirm that Sr$_{2}$CuO$_{3}$  may, nevertheless, be considered as 
an excellent realization of a 1D $S$=1/2 Heisenberg antiferromagnet.

We acknowledge useful discussions with X.~Zotos and F.~Naef. This work was financially 
supported in part by 
the Schweizerische Nationalfonds zur F\"{o}rderung der Wissenschaftlichen 
Forschung.


\begin{references}
\bibitem{Comment1} Note that the possibility of an enhanced 
    low-temperature phonon mean free path along the $b$ axis is not 
    taken into account in this calculation.
\end{references}
\end{document}